# *Ab initio* determination of ultrahigh thermal conductivity in ternary compounds

Huan Wu, Hang Fan, and Yongjie Hu [*]

*School of Engineering and Applied Science, University of California, Los Angeles (UCLA), Los Angeles, California 90095, USA*

Discovering new materials with ultrahigh thermal conductivity has been a critical research frontier and driven by many important technological applications ranging from thermal management to energy science. Here we have rigorously investigated the fundamental lattice vibrational spectra in ternary compounds and determined the thermal conductivity using a predictive *ab initio* approach. Phonon transport in B-*X*-C (*X* = N, P, As) groups is systematically quantified with different crystal structures and high-order anharmonicity involving a four-phonon process. Our calculation found an ultrahigh room-temperature thermal conductivity through strong carbon-carbon bonding up to 2100 W m$^{-1}$ K$^{-1}$ beyond most common materials and the recently discovered boron arsenide. This study provides fundamental insight into the atomistic design of thermal conductivity and opens up opportunities in new materials searching towards complicated compound structures.

With the shrinking device size and increasing power density, heat dissipation is becoming a critical technology challenge for modern electronics [1–3]. Discovering new materials with high thermal conductivity (HTC) is an emerging research frontier to tackle the thermal management issue [4–9]. The current industrial HTC standard is copper and silicon carbide with thermal conductivity ($\kappa$) around 400 W m$^{-1}$ K$^{-1}$. Conventional materials with conductivity beyond this standard are prototyped by single-element materials, i.e., diamond, graphene, and carbon nanotubes that all process a simple crystal structure. With the recent efforts in unveiling the atomistic origin using *ab initio* theory [10–23], exciting progress has been made in predicting HTC in new binary compounds. In particular, building on the atomistic theory prediction [16], boron arsenide and boron phosphide have been experimentally verified [5–8,24] with high $\kappa$ of 1300 and 500 W m$^{-1}$ K$^{-1}$, respectively, as a record high beyond common semiconductors. Meanwhile, intensive studies have been made to examine binary compounds almost over the entire elementary table; however, materials with conductivity over 400 W/mK have been very limited so far [25,26]. Now the question is whether such an HTC can exist in even complicated crystal structure, perhaps ternary compounds, which has not yet been explored. Here *ab initio* calculation is carefully conducted to investigate the fundamental vibrational spectra of compounds B-*X*-C (*X* = N, P, As) with high-order phonon anharmonicity involving a four-phonon process, and a group of HTC materials is identified up to 2100 W m$^{-1}$ K$^{-1}$ beyond most common materials.

We use a predictive *ab initio* approach to determine the thermal conductivity and quantify the fundamental vibrational spectra [10–23]. In general, heat transfer in semiconductors is primarily carried by phonons, i.e., the quantized modes of lattice vibrations [27]. Based on the phonon theory, the lattice thermal conductivity ($\kappa$) of these ternary compounds is given as a tensor matrix

$$\kappa^{\alpha\beta} = \frac{1}{N} \sum_\lambda C_\lambda v_\lambda^\alpha F_\lambda^\beta, \qquad (1)$$

where $\alpha$ and $\beta$ denote the crystal directions and $\lambda \equiv (\mathbf{q}, p)$ labels a phonon mode with wave vector $\mathbf{q}$ and polarization $p$. $C_\lambda$ and $v_\lambda^\alpha$ are the volumetric specific heat and the group velocity along $\alpha$ direction of phonon mode $\lambda$, respectively. $N$ is the number of $\mathbf{q}$ points in the mesh of the Brillouin zone. Importantly, $F_\lambda^\alpha$ measures the deviation of the nonequilibrium distribution function ($n_\lambda$) from equilibrium Bose-Einstein distribution ($n_\lambda^0$) under a given temperature gradient $\nabla T$, $n_\lambda = n_\lambda^0 + (-\partial n_\lambda^0/\partial T)\mathbf{F}_\lambda \cdot \nabla T$. $F_\lambda^\alpha$ has the dimension of length and can be regarded as effective phonon mean free path (MFP) along $\alpha$ direction. For most materials, $\nabla T$ does not drive $n_\lambda$ far from equilibrium, so the scattering rates $\tau_\lambda^{-1}$ for each individual phonon mode $\lambda$ can be calculated by keeping background phonons in equilibrium, and in this case $\mathbf{F}_\lambda = \tau_\lambda \cdot \mathbf{v}_\lambda$. However, for high-thermal-conductivity materials, $n_\lambda$ is usually driven far away from equilibrium by an applied $\nabla T$, so the deviation of all the phonon modes should be considered simultaneously by determining $\mathbf{F}_\lambda$ from the phonon Boltzmann transport equation (BTE) through self-consistent iteration [28]. In BTE, the phonon flux driven by the applied temperature gradient $\nabla T$ is balanced by phonon scattering,

$$\mathbf{v}_\lambda \cdot \nabla T \frac{\partial n_\lambda}{\partial T} = \left(\frac{\partial n_\lambda}{\partial t}\right)_{\text{scattering}}. \qquad (2)$$

The right side of Eq. (2) represents the phonon scatterings that alter $n_\lambda$, including, e.g., three-phonon scatterings, four-phonon scatterings, and isotope scatterings determined by quantum perturbation theory [29]. Note that this *ab initio* approach uses no adjustable parameters: the phonon frequencies and velocities are determined by diagonalizing the dynamic matrix, and the $\mathbf{F}_\lambda$ is calculated from the linearized BTE;

[*]Corresponding author: yhu@seas.ucla.edu

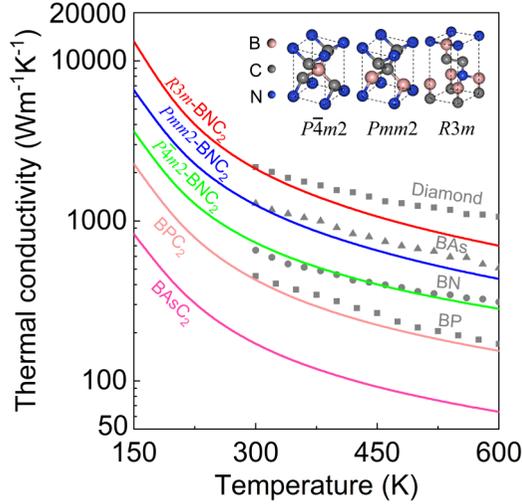

FIG. 1. Calculated lattice thermal conductivity ($\kappa$) vs temperature for ternary boron compounds B-$X$-C ($X$ = N, P, As). Our recently measured $\kappa$ values ($a$ axis) for binary boron compounds (BAs [6], BN [6], BP [5]) and diamond [6] are given for comparison. Inset shows the crystal structures for different phases of B-N-C.

the only inputs are the interatomic force constants that are obtained using density functional theory [30,31]. Such an approach has been verified in reliably calculating classical materials and predicting binary compounds that show good agreement with experiments [5–23]. More details regarding the calculation can be found in our recent work [6,21,22] and the Supplemental Material [32].

Here, *ab initio* theory is applied to systematically examine the ternary compounds B-$X$-C involving different crystal lattices [33], including the $R3m$ space group of rhombohedral lattice, the $Pmm2$ space group of primitive orthorhombic lattice, and the $P\bar{4}m2$ space group of primitive tetragonal lattice (inset, Fig. 1). These three crystal structures resemble that of diamond with strong atomic bonding, where each carbon atom is connected through four covalent bonds to form a pyramidal structure. For the $R3m$ phase, three bonds are with other carbon atoms and one bond is with a boron or nitrogen atom. For the $Pmm2$ phase, two bonds are with carbon atoms along the $a$ axis and the other two are alternating bonds with boron or nitrogen atoms along the $b$ axis. For the $P\bar{4}m2$ phase, two bonds are with boron atoms and the other two are with nitrogen atoms. Our calculation results of B-$X$-C with temperature dependence are plotted in Fig. 1; plotted together are our recent experimental measurements of the binary counterparts B-$X$ [5,6] and diamond [6]. In particular, $R3m$-BNC$_2$ shows an exceptionally high room-temperature $\kappa$ of 2100 Wm$^{-1}$K$^{-1}$ (Fig. 1), exceeding that of the recently reported BAs [6–8,16]. We note that no measurements of $\kappa$ for these ternary compounds have been reported to date, but crystals of BNC$_2$ have been synthesized at high pressure and high temperature [34–37] and shown with a high mechanical hardness [35,36].

The phonon band structures of these ternary compounds are calculated to show three acoustic and nine optic branches [along $\Gamma - X$, Fig. 2(a)]. From N to P to As, the frequency distribution range scales down because the increased average mass leads to large inertia to impede the atomic vibrations. Quantitatively, this leads to decreased acoustic phonon velocities and increased phonon population, resulting in increased phonon-phonon scattering. In general, thermal conductivity is governed by phonon scattering from the anharmonicity of the interatomic potential [27]. Note that all scatterings must satisfy both energy conservation and momentum conservation; on that perspective, with decreased phonon frequencies, it is easier to find available phonon modes that satisfy both conservation rules and thus increasing phonon scattering channels. To quantitatively illustrate that, we calculate the total three-phonon scattering phase space ($P_3^{\text{total}}$) [21], which quantifies the number of allowed three-phonon scattering channels,

$$P_3^{\text{total}} = \frac{1}{3Nm} \sum_\lambda \left(2P_{3\lambda}^{(+)} + P_{3\lambda}^{(-)}\right), \quad (3)$$

where $m$ is the number of phonon branches, and $N$ is the number of $\boldsymbol{q}$ points in the mesh of the Brillouin zone. $P_{3\lambda}^{(+)}$ and $P_{3\lambda}^{(-)}$ are the mode-dependent three-phonon scattering phase space for combination ($\omega_\lambda + \omega_{\lambda'} = \omega_{\lambda''}$) and splitting ($\omega_\lambda = \omega_{\lambda'} + \omega_{\lambda''}$) processes, respectively:

$$P_{3\lambda}^{(\pm)} = \frac{1}{Nm^2} \sum_{\lambda',\lambda''} \delta(\omega_\lambda \pm \omega_{\lambda'} - \omega_{\lambda''})\delta_{\boldsymbol{q}\pm\boldsymbol{q}',\boldsymbol{q}''+\boldsymbol{G}}. \quad (4)$$

Due to fact that $\delta(\beta\omega) = \frac{1}{\beta}\delta(\omega)$, where $\beta$ is the scaling factor of $\omega$, scaling down the phonon frequency could scale up the $P_3^{\text{total}}$. Actually, $P_3^{\text{total}}$ increases from 4.6 fs for BNC$_2$ towards 8.8 fs for BPC$_2$ and 11.8 fs for BAsC$_2$. The phonon scattering rates are calculated for the ternary compounds for comparison [Fig. 2(b)] and verify such a trend. In addition, the rapid increasing $\kappa$ of BXC$_2$ to exceed that of their binary compounds with decreased temperature, as observed in Fig. 1, is due to the suppression of acoustic-optic phonon scattering as optical phonons are frozen. In contrast, a large acoustic-optical band gap exists in binary boron compounds, suppressing the acoustic-optical scattering and giving a weaker temperature effect.

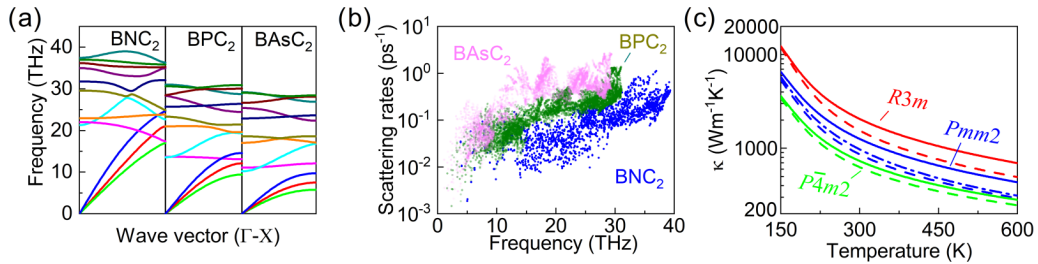

FIG. 2. Microscopic comparisons on thermal properties among different ternary boron compounds. (a) Calculated phonon band structures along $\Gamma - X$ and (b) phonon scattering rates at 300 K for BNC$_2$, BPC$_2$, and BAsC$_2$ of the $Pmm2$ phase. (c) Calculated $\kappa$ vs temperature for BNC$_2$ in $R3m$, $Pmm2$, and $P\bar{4}m2$ phases along the $a$ axis (solid line), $b$ axis (dash dot line), and $c$ axis (dash line). $\kappa$ along the $a$ axis and $b$ axis are degenerated for the $P\bar{4}m2$ and $R3m$ phases of BNC$_2$.

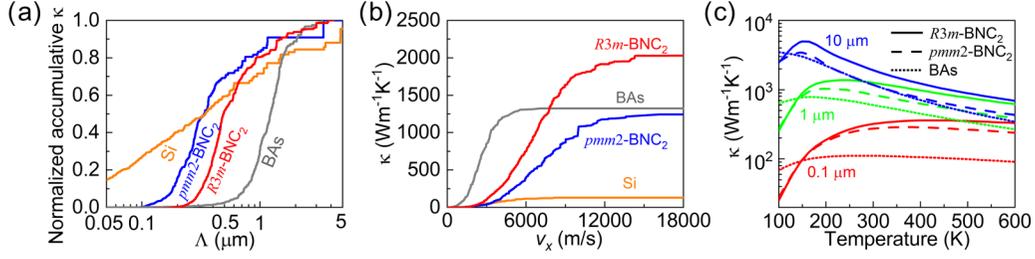

FIG. 3. Comparison on the contributions to $\kappa$ from spectral phonon mean free paths, phonon group velocities, and size effects. Calculated contributions to $\kappa$ from (a) phonon mean free paths and (b) phonon group velocity ($x$ component, $v_x$) for Si, BAs, $R3m$-BNC$_2$, and $Pmm2$-BNC$_2$ at 300 K. (c) Calculated $\kappa$ vs temperature for $R3m$-BNC$_2$, $Pmm2$-BNC$_2$, and BAs at different crystal sizes.

To investigate the effects on $\kappa$ from different atomic bonds, we compare the structural difference among the three different phases of B-N-C (inset, Fig. 1). In particular, the carbon atoms are distributed between B-C, $X$-C, and C-C bonds, among which the latter is expected to be the strongest [38]. The $R3m$ phase has more C-C bonds than the other two, while the $P\bar{4}m2$ phase does not have a C-C bond as the nitrogen atoms are switched with its neighboring carbon atoms from the $Pmm2$ phase. We calculated the thermal conductivity of all phases of BNC$_2$ and thermal conductivity along all directions are high despite an anisotropic structure [Fig. 2(c)]. Note that the atomic arrangement of $P\bar{4}m2$ and $R3m$ leads to a degeneracy symmetry and isotropic thermal conductivity along the $a$ and $b$ axis at 300 K. As expected, $\kappa$ of the $R3m$ phase is the highest, at 2100, 2100, and 1554 Wm$^{-1}$ K$^{-1}$ for the $a$, $b$, and $c$ axis, respectively. Also, $\kappa$ of the $Pmm2$ phase (1242, 958, and 846 Wm$^{-1}$ K$^{-1}$) is next to that of the $R3m$ phase but higher than that of the $P\bar{4}m2$ phase (723, 723, and 620 W/mK, respectively). We attribute the peak thermal conductivity to the aligned array of strong C-C bonds in $Pmm2$ that opens up an efficient heat conduction channel. Further, we calculated the lattice compatibility, defined as the number of atoms per unit volume, to be 1.719, 1.749, and 1.760×10$^{29}$ atoms/m$^3$ from $P\bar{4}m2$, $Pmm2$, to $R3m$ phase, indicating an increase in the interatomic bonding strength and thereby higher thermal conductivity.

To further investigate the difference between the binary and ternary boron compounds, we compare the phonon spectral information between BNC$_2$, BAs, and Si. The phonon MFP is the average distance that the phonons travel between two adjacent scatterings [5,39]. Here, an effective nondirectional MFP for each phonon mode $\Lambda_\lambda$ is defined as $\Lambda_\lambda \equiv \boldsymbol{F}_\lambda \cdot \boldsymbol{v}_\lambda / |\boldsymbol{v}_\lambda|$. The calculated cumulative MFP distribution is shown in Fig. 3(a), which quantifies the percentage contribution to thermal conductivity from the modes with MFP smaller than a certain value $\Lambda_\lambda$. In general, long MFPs are observed in high-thermal-conductivity materials such as diamond, graphene, BP, and BAs [5,6]. For example, phonons with long MFPs (1–10 $\mu$m) contribute to a very high portion of BAs's total thermal conductivity ($>$50%) [6]. In contrast, for normal semiconductors such as Si, phonon MFPs are distributed over a wider range (1 nm to 100 mm) [39]. Here, however, BNC$_2$, despite an ultrahigh $\kappa$ comparable with that of BAs, has abnormally short phonon MFP distribution, i.e., up to 80% of $\kappa$ is contributed by the phonons with MFP smaller than 0.5 $\mu$m. To understand the origin of ultrahigh $\kappa$ under very short phonon MFPs in BNC$_2$, we calculated the cumulative $\kappa$ as a function of phonon group velocity in Fig. 3(b) and found a big contrast in contributions. For BAs and Si, $\kappa$ is mainly contributed by the phonons with group velocity below 6000 m/s. But for both phases of BNC$_2$, $\kappa$ is largely contributed by phonons with velocity from 6000 m/s up to 18,000 m/s. This result implies that the origin of the ultrahigh thermal conductivity of BNC$_2$ comes from the large propagation velocity of the heat carriers as a result of small atomic mass and strong covalent bonding. It should be noted that with comparable ultrahigh bulk $\kappa$, the shorter MFPs in BNC$_2$ could benefit in size scaling for thermal management applications as long MFP phonons experience additional boundary scattering at limited sizes. A simplified estimation considering the boundary scattering rate ($\tau_\lambda^b$) for a device with the size of $L$ as $1/\tau_\lambda^b = |\boldsymbol{v}_\lambda|/L$ gives the size-dependent $\kappa$ plotted in Fig. 3(c). This shows that when the device size $L$ is scaled below 1 $\mu$m, BNC$_2$ enables more efficient heat dissipation than BAs. For example, when $L = 100$ nm, $\kappa$ of BNC$_2$ is more than twice that of BAs above 200 K.

Further, we emphasize the importance of high-order anharmonicity in considering ultrahigh thermal conductivity due to the recent advance in the field, and we have spent efforts to carefully perform the four-phonon calculations to ensure the prediction accuracy (Supplemental Material [32]). For many decades, thermal conductivity of solids was considered to be governed by the three-phonon scattering process, and the effects of four-phonon and higher-order processes were believed to be negligible by most past literature [25,27,40]. This is because the scattering probability of four-phonon process is in general expected to be low due to the momentum and energy conversation, as well as it requires a dramatically increased computational cost [18,40,41]. As a result, calculation of the four-phonon scattering process has only been carefully conducted until very recently [42]. Experimentally, recent study on the high-thermal-conductivity binary boron compound BAs has verified a very strong four-phonon scattering that significantly reduces $\kappa$ by about 40% from the three-phonon prediction [6]. Therefore calculation including the four-phonon processes has become critical for examining high-thermal-conductivity materials, despite its substantial computational challenge over the three-phonon process. Here we conducted four-phonon calculations and made the comparison with three-phonon-only results for the isotopically pure BAs and BNC$_2$. Our results [Fig. 4(a)] show that for BNC$_2$, four-phonon scattering has a much weaker effect (for example, with a reduction in $\kappa$ by 9% for $R3m$-BNC$_2$, i.e., from 2300 to 2100 W/mK) at room temperature and only up to 20% at high temperature (i.e., 600 K). Further, we analyze the frequency-specific contributions to develop a microscopic comparison. Figure 4(b) presents the $\kappa$ contributions from different phonon frequencies, with and without considering four-phonon scattering. Indeed, this verifies the weak four-phonon scattering in BNC$_2$ over the whole spectra. More quantitatively, we plot the mode-dependent three-phonon scattering rates ($\tau_3^{-1}$) and four-phonon scattering rates ($\tau_4^{-1}$) [Fig. 4(c)], which shows that that $\tau_3^{-1}$ is up to two orders larger than $\tau_4^{-1}$ for a large portion of phonon modes. The ratio $\tau_4^{-1}/\tau_3^{-1}$ of BNC$_2$ is much smaller than that of BAs [inset, Fig. 4(c)], verifying that in relation to three-phonon scattering, the four-phonon scattering is very weak for BNC$_2$, while it is much stronger in BAs. In addition, here we derived a four-phonon scattering phase space ($P_4^{\text{total}}$) to quantify the available four-phonon scattering channels that satisfy both momentum and energy conservation,

$$P_4^{\text{total}} = \frac{1}{7Nm} \sum_{\lambda} \left( 3P_{4\lambda}^{(++)} + 3P_{4\lambda}^{(+-)} + P_{4\lambda}^{(--)} \right), \quad (5)$$

$$P_{4\lambda}^{(\pm\pm)} = \frac{1}{N^2 m^3} \sum_{\lambda',\lambda'',\lambda'''} \delta(\omega_\lambda \pm \omega_{\lambda'} \pm \omega_{\lambda''} - \omega_{\lambda'''})$$
$$\times \delta_{q\pm q'\pm q'',q'''+G}, \quad (6)$$

where $P_{4\lambda}^{(++)}$, $P_{4\lambda}^{(+-)}$, and $P_{4\lambda}^{(--)}$ are mode-dependent four-phonon scattering phase spaces for combination ($\omega_\lambda + \omega_{\lambda'} + \omega_{\lambda''} = \omega_{\lambda'''}$), redistribution ($\omega_\lambda + \omega_{\lambda'} = \omega_{\lambda''} + \omega_{\lambda'''}$), and splitting ($\omega_\lambda = \omega_{\lambda'} + \omega_{\lambda''} + \omega_{\lambda'''}$) processes. The $P_4^{\text{total}}$ of BAs is 29 fs, while the $P_4^{\text{total}}$ of $R3m$-BNC$_2$ and $Pmm2$-BNC$_2$ are only 11.1 and 10.6 fs, respectively, indicating BNC$_2$ has much less available four-phonon scattering channels than BAs. These quantitative analyses have clearly illustrated the origin of HTC from the microscopic understanding of lattice vibrational spectra.

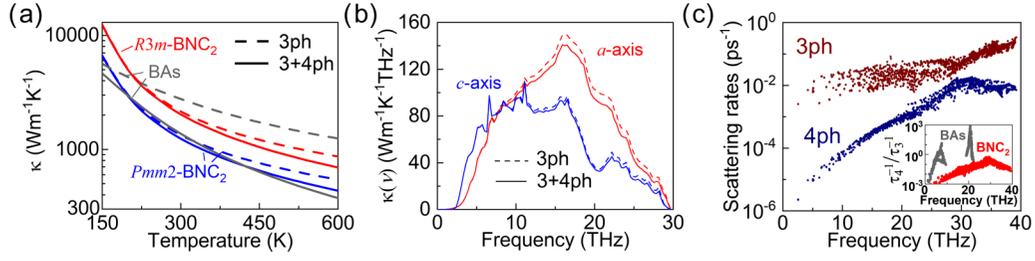

FIG. 4. Spectral contributions from four-phonon scattering vs three-phonon scattering. (a) Calculated $\kappa$ of BAs, $R3m$-BNC$_2$, and $Pmm2$-BNC$_2$, with the consideration of only the three-phonon process (3ph) vs the consideration of both three- and four-phonon scattering processes (3+4ph). (b) Calculated spectral contributions to thermal conductivity [$\kappa(\nu)$] of $R3m$-BNC$_2$ along the $a$ axis (red) and $c$ axis (blue), with the consideration of the three-phonon vs three- and four-phonon scattering processes. (c) Comparison on the spectral scattering rates from the three- and four-phonon processes, respectively. The figure inset shows the ratio of four-phonon to three-phonon scattering rates for $R3m$-BNC$_2$ vs BAs.

In summary, this work has identified a group of ternary compound materials with ultrahigh thermal conductivity beyond the state of the art. We have carefully investigated the phonon physics and high-order anharmonicity, including both three- and four-phonon scattering processes. The study represents significant progress in discovering new materials using *ab initio* prediction and provides important fundamental insight into the rational materials design for future thermal energy applications and new opportunities.

Y.H. acknowledges support from a CAREER Award from the National Science Foundation (NSF) under Grant No. DMR-1753393, a Young Investigator Award from the United States Air Force Office of Scientific Research under Grant No. FA9550-17-1-0149, an Alfred P. Sloan Research Fellowship under Grant No. FG-2019-11788, a PRF Doctoral New Investigator Award from the American Chemical Society under Grant No. 58206-DNI5, the Sustainable LA Grand Challenge, and the Anthony and Jeanne Pritzker Family Foundation. This work used computational and storage services associated with the Hoffman 2 Shared Cluster provided by the UCLA Institute for Digital Research and Education's Research Technology Group, and the Extreme Science and Engineering Discovery Environment (XSEDE), which is supported by National Science Foundation through Grant No. ACI-1548562.